\documentclass[conference,a4paper]{IEEEtran}
\IEEEoverridecommandlockouts
\usepackage{cite}
\usepackage{amsmath,amssymb,amsfonts}
\usepackage{algorithmic}
\usepackage[linesnumbered,ruled,vlined]{algorithm2e}
\usepackage{graphicx}
\usepackage{textcomp}
\usepackage{xcolor}
\usepackage{caption}
\usepackage{mathtools}

\usepackage{subcaption}
\def\BibTeX{{\rm B\kern-.05em{\sc i\kern-.025em b}\kern-.08em
    T\kern-.1667em\lower.7ex\hbox{E}\kern-.125emX}}
\begin{document}

\title{
Integration of SDN and Digital Twin for the Intelligent Detection of DoC Attacks in WRSNs}

\author{
	\IEEEauthorblockN{
		\normalsize Muhammad~Umar~Farooq~Qaisar$^{1}$, 
		\IEEEmembership{Member,~IEEE}, 
		Weijie~Yuan$^{2}$, \IEEEmembership{Senior Member,~IEEE}, \\
		Guangjie~Han$^{3}$, \IEEEmembership{Fellow,~IEEE},
		Adeel~Ahmed$^{4}$, \IEEEmembership{Graduate Student Member,~IEEE}, \\
		Chang~Liu$^{5}$, \IEEEmembership{Member,~IEEE}, and
		Md.~Jalil~Piran$^{6}$, \IEEEmembership{Senior Member,~IEEE}
	}
	
	\IEEEauthorblockA{
		$^{1}$\,\normalsize School of Computer Science, Northwestern Polytechnical University, China\\
		$^{2}$\,\normalsize School of System Design and Intelligent Manufacturing, Southern University of Science and Technology, China\\
		$^{3}$\,\normalsize Department of Internet of Things Engineering, Hohai University, China\\
		$^{4}$\,\normalsize School of Computer Science and Technology, University of Science and Technology of China, China\\
		$^{5}$\,\normalsize School of Information Engineering, Guangdong University of Technology, Guangzhou, China\\
		$^{6}$\,\normalsize Department of Computer Science and Engineering, Sejong University, Seoul 05006, South Korea\\
	}
}


\maketitle

\begin{abstract}
Wireless rechargeable sensor networks (WRSNs), supported by recent advancements in wireless power transfer (WPT) technology, hold significant potential for extending network lifetime. However, traditional approaches often prioritize scheduling algorithms and network optimization, overlooking the security risks associated with the charging process, which exposes the network to potential attacks. This paper addresses this gap by integrating Software-Defined Networking (SDN) and Digital Twin technologies for the intelligent detection of Denial of Charging (DoC) attacks in WRSNs. First, it leverages the flexibility and intelligent control of SDN, in combination with Digital Twin, to enhance real-time detection and mitigation of DoC attacks. Second, it employs four key metrics to detect such attacks including charging request patterns, energy consumption, behavioral and reputation scores, and charging behavior and efficiency. The numerical results demonstrate the superior performance of the proposed protocol in terms of energy usage efficiency, survival rate, detection rate, and travel distance. 
\end{abstract}


\section{Introduction}
Sixth-generation (6G) networks aim to support a wide range of applications that enhance connectivity and performance, particularly for emerging technologies such as digital twin networks, space-air-ground integrated networks, non-terrestrial networks, and others \cite{C0}. A key aspect of these advancements is the integration of Wireless Sensor Networks (WSNs), which are integral to the broader Internet of Things (IoT) framework \cite{C00}. The evolution of WSNs has become crucial in supporting 6G applications, including smart cities, industrial automation, military systems, environmental monitoring, and healthcare. These networks consist of small, battery-powered sensor nodes that monitor environmental conditions and transmit data to a central sink. However, the limited battery life of these nodes poses a significant challenge to network longevity, making energy consumption a critical concern. To address this, efficient energy management strategies such as duty cycling, adaptive power control, and energy-efficient routing are essential \cite{C1}. Additionally, incorporating rechargeable capabilities into Wireless Rechargeable Sensor Networks (WRSNs) can help mitigate energy depletion and extend network lifetimes \cite{C2}.

Despite extensive research on optimizing WRSN performance through charging scheduling and system improvements, security remains a relatively underexplored concern. Compromised sensor nodes can undermine both network integrity and energy management, leaving WRSNs vulnerable to emerging threats and attacks. This underscores the need for robust security mechanisms to protect WRSNs from potential risks.


One such critical security threat in WRSNs is the \textit{Denial of Charging} (DoC) attack, where malicious entities prevent legitimate sensor nodes from accessing charging, disrupting network operation. Attackers, which may include compromised nodes, malicious vehicles, or external adversaries, are capable of overwhelming the charging infrastructure with fake requests, jamming communication channels, or physically blocking Mobile Charging Vehicles (MCVs). This denial of access reduces sensor node lifetime and compromises overall network performance, potentially leading to network failure. Addressing DoC attacks is essential for maintaining the integrity and longevity of WRSNs \cite{C3}.

To tackle these challenges, \textit{Digital Twin} (DT) technology and \textit{Software-Defined Networking} (SDN) are emerging as promising solutions for enhancing the management, security, and optimization of WRSNs. A DT is a virtual representation of the actual WRSN, enabling real-time monitoring, simulation, and optimization of network operations. By continuously updating the digital model of the network, DT can help predict and prevent energy depletion issues, identify potential security vulnerabilities, and optimize charging schedules and node behaviors. It also allows for efficient anomaly detection, such as identifying unusual patterns indicative of a DoC attack, and enables predictive maintenance, which can help extend the life of the network \cite{C4}\cite{C5}.

Incorporating SDN into WRSNs further strengthens the network's management capabilities. SDN separates the control plane from the data plane, providing centralized control over network operations, which allows for dynamic and flexible energy management and routing protocols \cite{C6}. SDN can be used to monitor and control access to charging, ensuring that only authorized nodes can access it and mitigating the risk of DoC attacks. Additionally, SDN can facilitate adaptive power control and load balancing across the network, ensuring optimal energy distribution and improving overall network performance. When integrated with DT, SDN can enable intelligent decision-making by using real-time data and simulations to optimize both energy consumption and network security.

The aforementioned studies provide strong motivation to address the issue of DoC attacks and to develop a novel detection method for these attacks in WRSNs. Therefore, this paper proposes an integration of SDN and Digital Twin for the intelligent detection of DoC attacks in WRSNs. The contributions of our paper lie in several unique aspects. Firstly, it integrates SDN and DT technologies to enhance the detection and mitigation of DoC attacks in WRSNs. Specifically, the DT model is used to simulate the replica network behavior of the WRSN, while the SDN controls both the replica network and the WRSN for flexible management. Secondly, it employs four key metrics to detect DoC attacks at the DT layer: charging request patterns, energy consumption and residual energy, behavioral and reputation scores, and charging behavior and efficiency. These metrics are combined into a single maliciousness score, which aggregates the individual attack indicators into one value to assess the overall maliciousness of a node. This score enables the SDN controller to prioritize which nodes to charge and reconfigure the network to avoid further disruptions. The unique aspects and contributions of our approach make it an innovative and effective solution to the identified issues.

\section{Related Work}
This section presents a review of several studies addressing security challenges in WRSNs, which are closely related to our research.

In the field of network security, much of the research has focused on addressing the issue of electromagnetic radiation in charging applications \cite{C7}. Specifically, the authors introduced the concept of "safe charging" and aimed to maximize the charging efficiency of devices by adjusting the power levels of mobile chargers. However, their work is primarily concerned with mitigating the harmful effects of electromagnetic radiation on human health and does not address the potential security risks associated with charging attacks or propose corresponding countermeasures. In \cite{C3}, the authors introduced a DoC attack targeting WRSNs. In this type of attack, malicious nodes generate fake charging requests, leading to the depletion of energy in legitimate nodes and causing a degradation in the overall network performance. In \cite{C8}, the authors introduced a charging coordination approach that integrates blockchain and fog computing technologies. This approach employs fog computing nodes to determine charging schedules while ensuring secure communication through the use of mutual authentication mechanisms. In \cite{C9}, the authors explore the development of a DoC attack targeting WRSNs, with the goal of disrupting network operations by exploiting a malicious mobile charging vehicle. They formalize the problem maximization of destructiveness (MAD) and proposes a new attack method called MDoC, which offers performance guarantees for its effectiveness. The MDoC attack consists of two phases: the first phase prompts nodes to generate a number of charging requests, creating a request explosion, while the second phase identifies the longest charging path, leading to the depletion of energy in nodes and causing them to fail. In addition, various attack strategies in WSNs \cite{C10}, such as Denial of Service (DoS) attacks \cite{C11}, Distributed Denial of Service (DDoS) attacks \cite{C12}, and node capture attacks \cite{C13}, offer valuable insights for developing DoC attacks and merit consideration. While DoC attacks have not been explicitly discussed in existing literature, they share common characteristics with DDoS attacks, as both aim to disrupt network functionality with comparable objectives.

Since the DoC attack discussed in this paper is based on an on-demand charging architecture, several state-of-the-art scheduling methods are reviewed and summarized for reference. In \cite{C14}, the authors proposed a probabilistic on-demand charging scheduling scheme for ISAC-assisted WRSNs, which efficiently prioritized charging nodes to optimize charging efficiency. In \cite{C15}, the authors designed an on-demand recharging and data collection for WRSNs, which aims to serve the ideal data delivery and optimal charging scheduling. In \cite{C16}, the authors introduced a mixed-priority charging scheduling algorithm (mTS) that considers both temporal and spatial factors in an integrated manner. 

Despite significant advancements in research on WRSNs, the impact of potential attacks has largely been overlooked. This study addresses these threats by introducing a novel DoC attack detection and mitigation approach, which targets disruptions to network functionality within an on-demand charging architecture.

\section{The Proposed Protocol} \label{TPP}
This paper proposes an integration of SDN and Digital Twin for the intelligent detection of DoC attacks in WRSNs. It outlines comprehensive strategies to enhance the detection and mitigation of such attacks by leveraging the Digital Twin model alongside SDN's flexible management capabilities. The SDN controller computes four key attributes to identify potential attacks within the Digital Twin model, while simultaneously updating the charging queue in the actual WRSN to prevent disruptions in the charging process. Additionally, by utilizing real-time data from the Digital Twin, the SDN controller can detect anomalies and make immediate, adaptive decisions. This real-time decision-making strengthens the resilience of WRSNs against evolving security threats.

\subsection{Network Model and Initialization} \label{NMA}
In this study, we consider a standard wireless rechargeable sensor network (WRSN) consisting of a set of randomly deployed sensor nodes, $\mathcal{N} = \{{n_{0}, n_{1}, n_{2},..., n_{r}}\}$, and a set of mobile charging vehicles (MCVs), $\mathcal{V} = \{{Cv_{1}, Cv_{2}, Cv_{3},..., Cv_{m}}\}$, where $m = |\mathcal{V}|$ represents the number of MCVs in the network. When the energy level of a sensor node ($n_{i}$) drops below a predefined threshold $\mathcal{E}{th}$, it sends a charging request to the sink node ($n_{b}$), which is located at the center of the region and also serves as the depot for the MCVs. We assume that the MCVs follow the charging strategy as detailed in our previous work \cite{C17}.
The initial positions of the MCVs are calculated using the method described in \cite{C18}, with the coordinates $(x_{i}, y_{i})$ of the MCVs determined in $k$ different regions as follows.

\begin{equation} \label{MCV1}
	x_{i} = \frac{\mathcal{C}_{c}}{2}\cos{\left(\frac{\pi}{m}(2j-1)\right)},
\end{equation}

\begin{equation} \label{MCV2}
	y_{i} = \frac{\mathcal{C}_{c}}{2}\sin{\left(\frac{\pi}{m}(2j-1)\right)},
\end{equation}

where $\mathcal{C}_{c}$ is the radius of the circumscribed circle around the two-dimensional region. Each MCV recharges one sensor node at a time, and it is assumed that MCVs have a significantly higher energy capacity than the sensor nodes. 

\begin{figure}[t]
	\includegraphics[width=\linewidth,height=8cm]{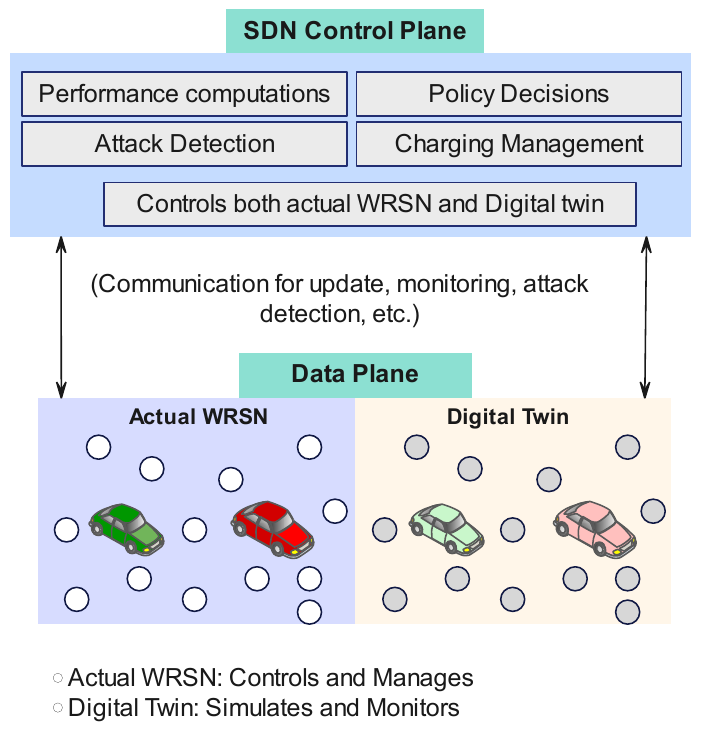}
	\centering
	\caption{SDN-controlled WRSN and Digital Twin architecture}
	\captionsetup{justification=centering}
	\label{fig:SWDA}
\end{figure}

We employ a Digital Twin (DT) as a real-time virtual replica of the WRSN, enabling continuous simulation and monitoring. The SDN controller integrates both the physical WRSN and DT, using real-time data to compute and update key attributes, including the four scores. These scores assess network status, detect anomalies, and predict security threats. Leveraging the DT’s simulation, the SDN controller dynamically adjusts the charging queue and makes informed decisions. This integration improves detection and response to compromised nodes, refines mitigation strategies, and optimizes maliciousness scoring thresholds. The coordinated approach ensures proactive, data-driven decisions, allowing the network to adapt to emerging threats while improving performance. Fig. \ref{fig:SWDA} illustrates the SDN control plane managing both the WRSN and DT for monitoring, updates, and attack detection. Table \ref{Notations} lists all notations used in this work.

\begin{table}[t]
	\centering
	\caption{Notations}
	\label{Notations}
	\begin{tabular}
		{cp{7cm}}
		\hline
		\textbf{Notation} & \textbf{Definition} \\
		\hline
		$\mathcal{N}$  & $\mathcal{N} = \{{n_{0}, n_{1}, n_{2},..., n_{r}}\}; n_{i} \in \mathcal{N}$ is a sensor node$; r$ is the size of $\mathcal{N}$.\\
		$\mathcal{N}_{i}$ & The neighboring nodes of $n_{i}$; $r_{i}$ is the size of $\mathcal{N}_{i}$. \\
		$\mathcal{V}$  & $\mathcal{V} = \{{Cv_{1}, Cv_{2}, Cv_{3}, ..., Cv_{m}}\}; Cv_{j} \in \mathcal{V}$ is a mobile charging vehicle$; m$ is the size of $\mathcal{V}$.\\
  	    	$n_{s}$ & The source node.   \\
		$n_{b}$ & The sink.   \\
		$\mathcal{R}_{s}$ & The sensing range of node $n_{i}$. \\
		$\mathcal{R}_{c}$ & The communication range of node $n_{i}$. \\
            $\mathcal{E}_{th}$ & The residual energy threshold of node $n_{i}$. \\
            $\mathcal{CE}_{th}$ & The minimum working energy threshold of MCV $Cv_{j}$. \\
		$C_i(t)$ & The charging request patterns of a node $n_{i}$. \\
		$E_i(t)$ & energy consumption and residual energy
of node $n_{i}$. \\
		$ R_i(t)$ & The behavioral and reputation scores of node $n_{i}$. \\
        $\eta_i(t)$ & The charging behavior and efficiency of node $n_{i}$. \\
	\end{tabular}
\end{table}

\subsection{Attack Detection Strategy}
This section outlines the attack detection strategy, which includes four key attributes and their probabilistic models to analyze the anomalous behavior of charging request nodes. The attributes are charging request patterns, energy consumption and residual energy, behavioral and reputation scores, and charging behavior and efficiency. Each attribute score is aggregated into a combined maliciousness score using a weighted sum approach to determine the threshold for detecting the DoC attack.

\subsubsection{Probabilistic Model for Charging Request Patterns}
We define the charging request rate \( C_i(t) \) for node \( n_i \) at time \( t \) as a random variable with a distribution representing normal charging behavior. The number of charging requests follows a Poisson distribution, which is often used to model events occurring at a constant rate given as follows.
\begin{equation} \label{CR}
C_i(t) \sim \text{Poisson}(\lambda_i),
\end{equation}
where \( \lambda_i \) is the expected charging request rate for node \( n_i \). For a node \( n_i \), the probability of an observed charging request \( C_i(t) \) at time \( t \) is given as;
\begin{equation} \label{CR1}
P(C_i(t) \mid \lambda_i) = \frac{\lambda_i^{C_i(t)} e^{-\lambda_i}}{C_i(t)!}.
\end{equation}
If \( C_i(t) \) significantly deviates from \( \lambda_i \), it indicates a potential anomaly (i.e., DoC attack), where the maliciousness score for the node \( n_i \) at time \( t \) is given as;
\begin{equation} \label{CR2}
\mathcal{M}_{C_i}(t) = 1 - P(C_i(t) \mid \lambda_i).
\end{equation}
A higher value indicates a higher probability of malicious behavior.

\subsubsection{Probabilistic Model for Energy Consumption}
The energy consumption pattern of a node can also be modeled probabilistically. The energy consumption follows a normal distribution with mean \( \mu_i(t) \) and variance \( \sigma_i^2(t) \), representing typical energy usage is:
\begin{equation} \label{EC}
E_i(t) \sim N(\mu_i(t), \sigma_i^2(t)),
\end{equation}
where \( \mu_i(t) \) and \( \sigma_i^2(t) \) are the mean and variance of the energy consumption at time \( t \). The probability of the observed energy consumption \( E_i(t) \) is given as follows.
\begin{equation} \label{EC1}
\begin{split}
P(E_i(t) \mid \mu_i(t), \sigma_i^2(t)) &= \frac{1}{\sqrt{2\pi \sigma_i^2(t)}} \\
& \quad \exp\left( -\frac{(E_i(t) - \mu_i(t))^2}{2 \sigma_i^2(t)} \right)
\end{split}
\end{equation}
If the energy consumption $E_i$(t) deviates significantly from this expected distribution, the maliciousness score for the node \( n_i \) at time \( t \) is given as follows.
\begin{equation} \label{EC2}
\mathcal{M}_{E_i}(t) = 1 - P(E_i(t) \mid \mu_i(t), \sigma_i^2(t)).
\end{equation}
A high score indicates a potential DoC attack if the node does not deplete energy or behaves differently from normal nodes.

\subsubsection{Probabilistic Model for Behavioral and Reputation Scores}
The reputation score \( R_i(t) \) of a node is modeled as a Markov chain process, where the current reputation score depends on the previous state (previous reputation), the behavior of the node, and feedback from other nodes. The state transition matrix \( P \) for the reputation process is given as:
\begin{equation} \label{BRS}
P(R_i(t+1) \mid R_i(t)) = P(R_i(t+1) = r \mid R_i(t) = r').
\end{equation}
where \( r' \) represents the previous reputation state at time \( t \) and \( r \) represents the future reputation state at time \( t+1 \).

The reputation score follows a Beta distribution characterized by parameters $\alpha$ and $\beta$, reflecting the updating mechanism described as follows.
\begin{equation} \label{BRS1}
R_i(t) \sim \text{B}(\alpha_i(t), \beta_i(t)),
\end{equation}
where $\alpha_i(t)$ and $\beta_i(t)$ represent the parameters updated based on node interactions. The probability of observing the reputation score at time \( t \) is given as:
\begin{equation} \label{BRS2}
P(R_i(t) \mid \alpha_i(t), \beta_i(t)) = \frac{R_i(t)^{\alpha_i(t) - 1}(1 - R_i(t))^{\beta_i(t) - 1}}{B(\alpha_i(t), \beta_i(t))},
\end{equation}
where $B(\alpha_i(t), \beta_i(t))$ is the Beta function. A deviation in the reputation score from its expected distribution indicates abnormal behavior, which increases the maliciousness score, as shown follows.
\begin{equation} \label{BRS3}
\mathcal{M}_{R_i}(t) = 1 - P(R_i(t) \mid \alpha_i(t), \beta_i(t)).
\end{equation}

\subsubsection{Probabilistic Model for Charging Behavior and Efficiency}
A probabilistic model for charging behavior and efficiency aims to capture how efficiently a node engages in charging requests and the effectiveness of energy transfer. This metric can model charging efficiency, the charging duration, and how these factors align with expected patterns under normal operation.
We define the charging efficiency $\eta_i(t)$ as the ratio of the energy successfully received by node $i$ over the total energy sent by the MCV, as shown as follows.
\begin{equation} \label{CBE}
\eta_i(t) = \frac{E_{\text{received},i}(t)}{E_{\text{sent},i}(t)}.
\end{equation}
This efficiency is modeled as a random variable $\eta_i(t)$ with a probabilistic distribution (normal distribution), based on expected charging behavior.
\begin{equation} \label{CBE1}
\eta_i(t) \sim N\left(\mu_i(t), \sigma_i^2(t)\right),
\end{equation}
where $\mu_i(t)$ is the expected charging efficiency, and $\sigma_i^2(t)$ is the variance. The probability of observing the actual charging efficiency is given follows.
\begin{equation} \label{CBE2}
\begin{split}
P\left(\eta_i(t) \mid \mu_i(t), \sigma_i^2(t)\right) &= \frac{1}{\sqrt{2\pi \sigma_i^2(t)}} \\
& \quad \exp\left( -\frac{\left(\eta_i(t) - \mu_i(t)\right)^2}{2 \sigma_i^2(t)} \right).
\end{split}
\end{equation}
If the charging efficiency deviates significantly from the expected behavior, it may indicate that a node is purposefully disrupting the charging process, an indication of a potential DoC attack.
The maliciousness score for this metric at time $t$ is computed as follows.
\begin{equation} \label{CBE3}
\mathcal{M}_{{\eta}_i}(t) = 1 - P\left(\eta_i(t) \mid \mu_i(t), \sigma_i^2(t)\right).
\end{equation}

A higher value indicates an abnormal or malicious charging behavior.

\subsubsection{Combined Maliciousness Score}
Now, we have computed the four metrics, each associated with a maliciousness score. Since these metrics represent different aspects of node behavior, we combine their individual maliciousness scores into an overall Combined Maliciousness Score using a weighted sum approach, as shown as follows.
\begin{equation} \label{CMS}
\begin{split}
\mathcal{M}_i(t) &= \omega_C \cdot \mathcal{M}_{C_i}(t) \\ & + \omega_E \cdot \mathcal{M}_{E_i}(t) \\ &+ \omega_R \cdot \mathcal{M}_{R_i}(t) \\ &+ \omega_{\eta} \cdot \mathcal{M}_{{\eta}_i}(t)
\end{split}
\end{equation}

Where $\omega_C, \omega_E, \omega_R, \omega_{\eta}$ are the weights assigned to each metric, reflecting their importance in detecting DoC or malicious behavior. The final score, $\text{Maliciousness}(i,t)$, is used to assess whether a node $n_i$ is behaving maliciously at time $t$.

\textbf{Threshold for DoC Attack Detection}: Once the Combined Maliciousness Score is obtained, we apply a threshold to determine whether a node is performing a DoC attack, as shown below.
\begin{equation} \label{CMS1}
DoC_{detect}(i,t) =
\begin{cases}
1 & \text{if } \mathcal{M}_i(t) > \theta_{\text{DoC}} \\
0 & \text{otherwise}
\end{cases}
\end{equation}
where $\theta_{\text{DoC}}$ is a threshold value, which can be set based on network's tolerance for malicious behavior. If the score exceeds this threshold, the node is flagged as potentially malicious DoC attack.

\subsection{SDN Controller Monitoring and Management of WRSN}
The SDN controller plays a critical role in managing both the virtual Digital Twin network and the actual WRSN. It simulates the virtual network's performance using four key attributes, providing a comprehensive view of node behavior and charging efficiency. This enables the SDN controller to monitor the network and calculate the Combined Maliciousness Score to detect and prevent potential DoC attacks. Upon detection, the SDN controller updates the charging queue of MCVs in the actual WRSN in real time, prioritizing normal nodes in need of urgent charging based on the criticality of their residual energy, while removing nodes with high maliciousness scores to prevent disruptions from potential DoC attacks as described in Algorithm \ref{SDW}. Through this adaptive mechanism, the SDN controller ensures continuous network optimization and resilience, leveraging the Digital Twin for simulation and the actual WRSN for real-time management.

\begin{algorithm}[t] 	
        \SetAlgoLined
        \caption{SDN and Digital Twin-Powered Autonomous WRSN Management}
        \label{SDW}

        Initialize virtual and actual WRSN networks.
        
        Continuously monitor and compute metrics.
	
        \textbf{function} Attack\_Detection
        
	{			
		\ForEach{node $n_{i}$ $\in$ $\mathcal{N}$}
		{				
			\textbf{Compute} \textit{Charging Request Patterns} \( C_i(t) \) \eqref{CR2} \\
			\textbf{Compute} \textit{Energy Consumption} \( E_i(t) \) \eqref{EC2} \\
			\textbf{Compute} \textit{Behavioral and Reputation Scores} \( R_i(t) \) \eqref{BRS3} \\
			\textbf{Compute} \textit{Charging Behavior and Efficiency} \( \eta_i(t) \) \eqref{CBE3} \\
			\textbf{Compute} \textit{Combined Maliciousness Score} \(\mathcal{M}_i(t) \) \eqref{CMS} \\
                \textbf{Compute} \textit{Threshold for DoC Attack Detection} \( DoC_{detect}(i,t) \) \eqref{CMS1} \\
		}
		
            Update charging queue of MCVs in actual WRSN.
            
            \ForEach{node $Cv_{j}$ $\in$ $\mathcal{V}$}
		{
                Identify nodes with high maliciousness scores or low residual energy.
                
                Update charging priorities based on \( DoC_{detect}(i,t) \) and energy requirements.
                
            }
            
	}
    
	\textbf{end function}
    
\end{algorithm}

\section{Performance Evaluation and Discussion}
In this section, we present the numerical results to evaluate the performance of the proposed protocol. The simulation is conducted in a square-shaped monitoring area with randomly deployed sensor nodes, while the sink node is positioned at the center. The sink serves as the central point for data collection and acts as the depot for MCVs to recharge their batteries \cite{C19}. For simplicity, the SDN controller is collocated with the sink node, functioning as an independent external device responsible for managing and controlling both the actual and virtual networks, as well as updating the charging queue priorities for MCVs. The simulation results are based on theoretical models and assume ideal conditions (e.g., no environmental interference or hardware failures).

The number of sensor nodes varies from $100$ and $500$, each with a communication range of $50$ meters and a sensing range of $25$ meters. The sensor nodes have a battery capacity of $0.5$ J. Each MCV is equipped with a battery capacity of $10$ kJ and a charging rate of $0.05$ J/s. The MCVs operate at an initial speed of $5$ m/s, with an energy consumption rate of $5$ J/m for travel. Additionally, the random waypoint mobility model is used to guide the movement of the MCVs.

\subsection{Results}
To the best of our knowledge, no prior work has addressed DoC attack detection. We evaluate our protocol by comparing its performance across different attack intensities, which are defined as Low Attack Intensity (LAI) ranging from 5\% to 10\%, Moderate Attack Intensity (MAI) from 20\% to 30\%, and High Attack Intensity (HAI) from 40\% to 50\%. The evaluation is conducted using the following performance metrics: i) \textbf{Energy Usage Efficiency} (\%): It is defined as the proportion of the total energy delivered to the sensor nodes relative to the total energy sent from the sink to the MCVs. ii) \textbf{Survival Rate} (\%): It is defined as the percentage of sensor nodes that remain operational compared to the total number of nodes in the network. iii) \textbf{Detection Rate} (\%): The ratio of malicious nodes accurately identified as attacks in the network compared to the total number of malicious attacks. iv) \textbf{Travel Distance} (m): It refers to the total distance traveled by the MCV during one complete charging cycle.

\begin{figure*}[t]
\centering
\subfloat[Energy Usage Efficiency]{\label{fig: EUENN}\includegraphics[width=.23\linewidth, height=3.4cm]{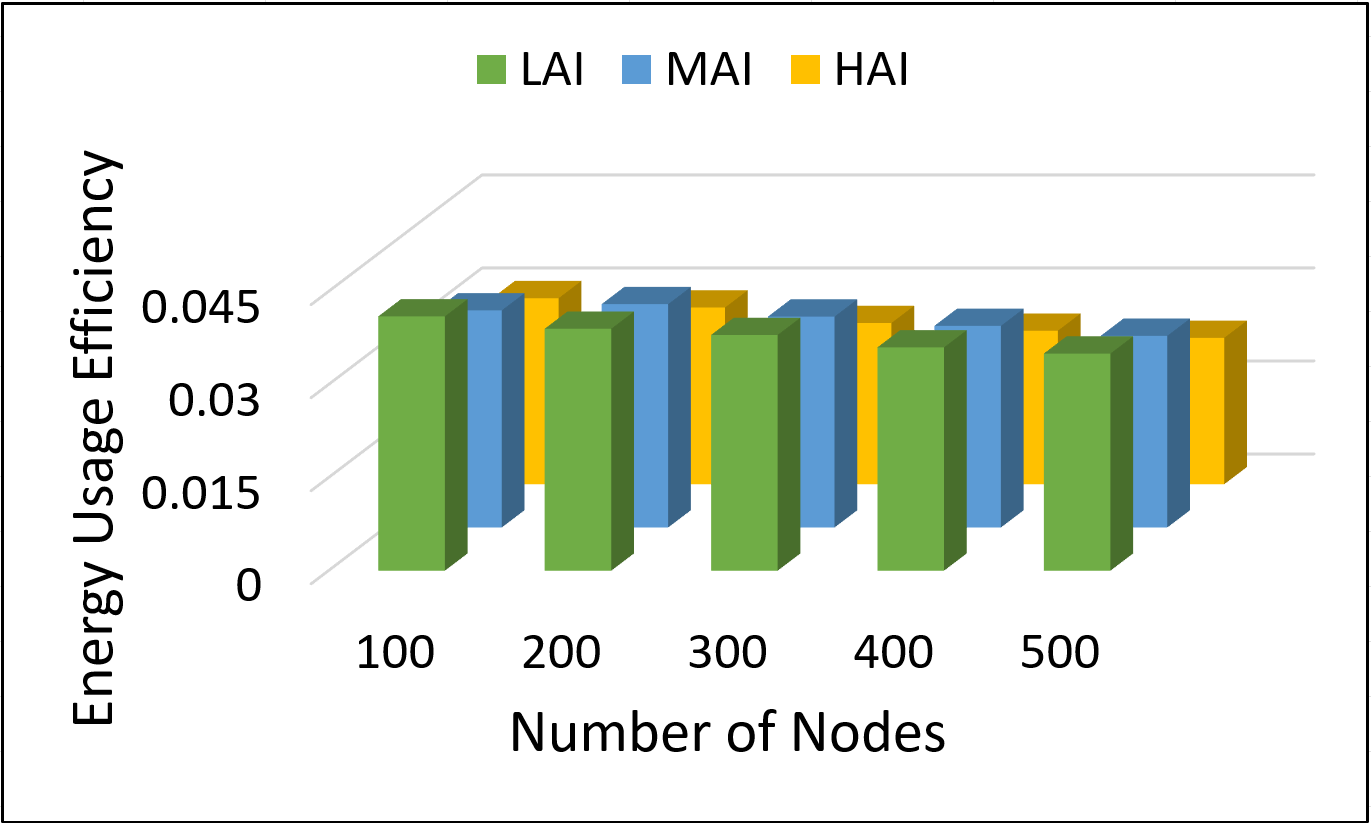}}\quad
\subfloat[Survival Rate]{\label{fig: SRNN}\includegraphics[width=.23\linewidth, height=3.4cm]{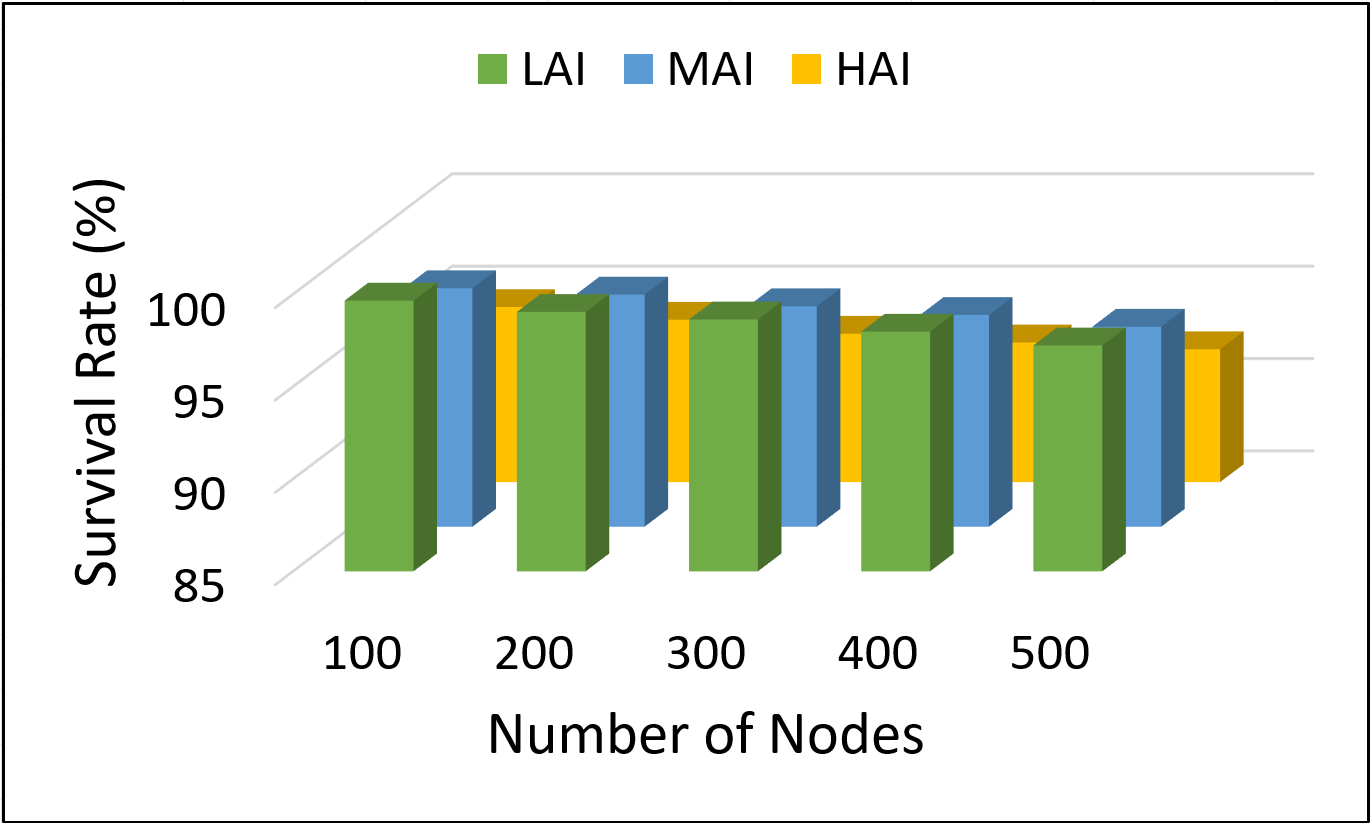}}\quad
\subfloat[Detection Rate]{\label{fig: DRNN}\includegraphics[width=.23\linewidth, height=3.4cm]{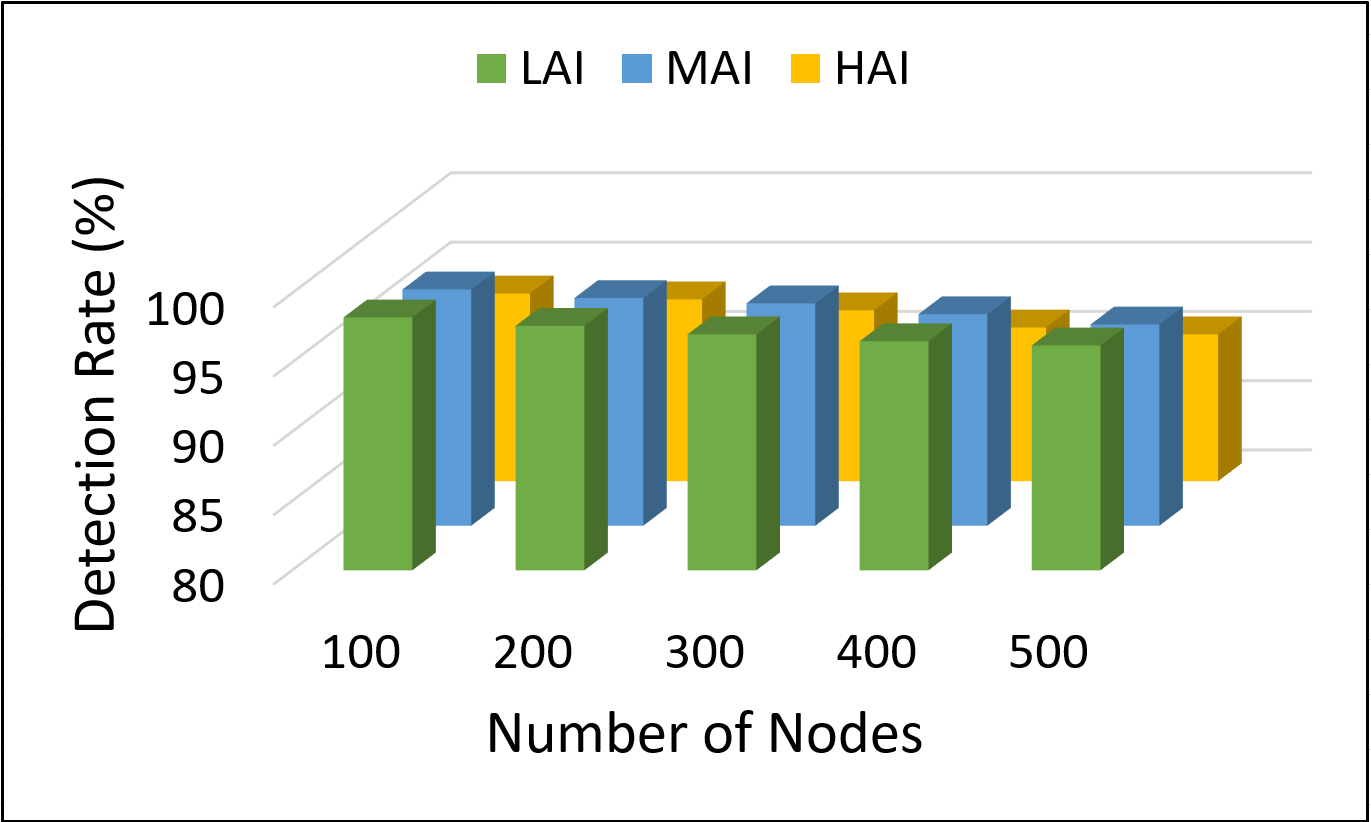}}\quad
\subfloat[Travel Distance]{\label{fig: TDNN}\includegraphics[width=.23\linewidth, height=3.4cm]{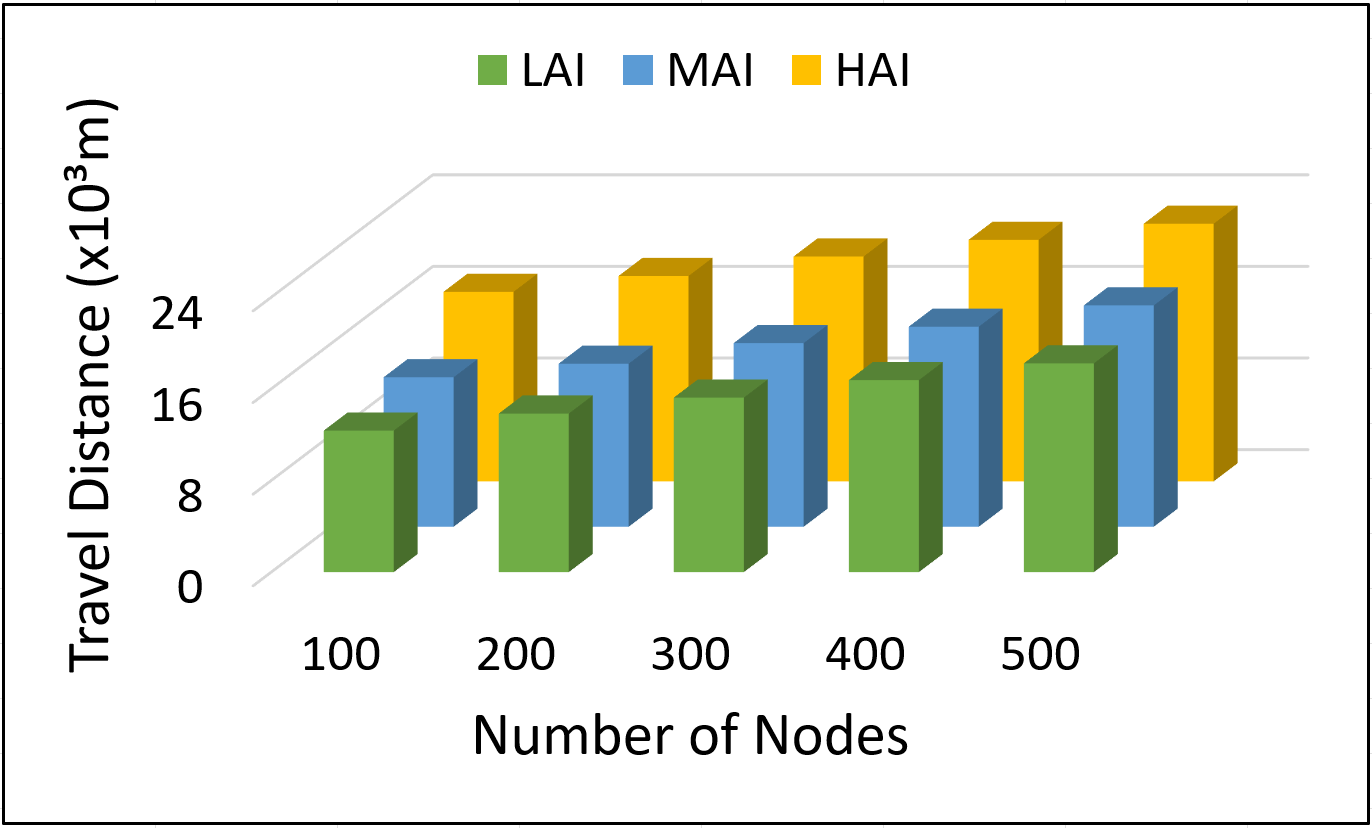}}
\caption{Performance over number of nodes}
\label{fig: PNN}
\end{figure*}

Fig. \ref{fig: EUENN} shows the energy usage efficiency, which gradually decreases as the number of nodes increases. It performs better across different attack intensities because SDN enables dynamic reconfiguration of the charging process, while Digital Twin provides real-time simulations to predict and detect DoC attacks. By promptly identifying and removing compromised nodes from the charging queue, the protocol prevents malicious nodes from draining energy resources, ensuring that legitimate nodes are efficiently charged without disruption.

Figure \ref{fig: SRNN} presents the survival rate, which gradually decreases as the number of nodes increases. It performs better in varying attack intensities because SDN's real time monitoring and control, in combination with Digital Twin’s predictive capabilities, allow for immediate identification and isolation of malicious nodes. This exclusion prioritizes legitimate nodes for charging, optimizing the survival rate even under high-intensity DoC attacks.

Figure \ref{fig: DRNN} shows the acceptable detection rate with the proposed protocol. By calculating the combined maliciousness score based on probabilistic models, the protocol enables fast and accurate detection of DoC attacks. It performs better across varying attack intensities because it can instantly identify disruptions caused by malicious nodes, preventing further impact on the network's performance. 

Figure \ref{fig: TDNN} illustrates the total travel distance of MCVs, which increases as the number of nodes increases. Our protocol controls the MCV travel distance by leveraging SDN's ability to prioritize and adjust charging schedules dynamically. With the help of Digital Twin simulations, the system identifies and removes malicious nodes from the charging queue in real time, reducing unnecessary detours and optimizing MCV routes. This results in more efficient network operation and shorter travel distances, even under high-intensity attacks.

\section{Conclusion}
This paper addressed a critical gap in the security of WRSNs, particularly in the context of DoC attacks. By integrating SDN and Digital Twin technologies, the proposed protocol enhances the real-time detection and mitigation of such attacks. The use of SDN enables flexible and intelligent control, while the Digital Twin facilitates accurate simulations to detect anomalies in network behavior. Four key metrics including charging request patterns, energy consumption, behavioral and reputation scores, and charging behavior were utilized in combined maliciousness score metric to detect potential DoC attacks, further strengthening the network's defense. The numerical results demonstrated the effectiveness of the proposed protocol, showing significant improvements in energy usage efficiency, survival rate, detection rate, and travel distance. This approach provide a robust solution for securing WRSNs and ensures the network’s resilience against attacks, ultimately contributing to the reliable operation of future IoT.

\bibliographystyle{ieeetr}
{\footnotesize
	\bibliography{sample-base}}

\end{document}